\documentstyle[11pt,aip,equations]{article}

\def\eqref#1{Eq.~(\ref{#1})}

\def\figref#1{Fig.~\ref{#1}}
\def\Figref#1{Figure~\ref{#1}}
\def\refcite#1{Ref.~\citenum{#1}}

\begin{document}

\begin{center}
{\large\bf Dependence of extensive chaos on the spatial
correlation length}\\
\ \\
{\bf David~A.\ Egolf$^\dagger$ and Henry S.~Greenside$^{\ast\dagger}$}\\
Department of Physics, Duke University,
Durham, North Carolina, 27708-0305 USA\\
$^\ast$ Also the Department of Computer Science, Duke University\\
$^\dagger$ Also the Center for Nonlinear and Complex Studies, Duke University\\
September 30, 1993\\
E-mail: dae@phy.duke.edu and hsg@cs.duke.edu
\end{center}

\ \hrulefill\   % horizontal line

{
\bf
We consider spatiotemporal chaotic systems
\cite{Ahlers85,Gollub90,Ouyang91,Arecchi92,Cross93} for
which spatial correlation functions decay substantially
over a length scale~$\xi$ (the spatial correlation
length) that is small compared to the system size~$L$.
Numerical simulations
\cite{Manneville85,Grassberger89,Sirovich90} suggest
that such systems generally will be {\em extensive},
with the fractal dimension~$D$ growing in proportion to
the system volume for sufficiently large systems ($L
\gg \xi$).  Intuitively, extensive chaos arises because
of spatial disorder. Subsystems that are sufficiently
separated in space should be uncorrelated and so
contribute to the fractal dimension in proportion to
their number.  We report here the first numerical
calculation that examines quantitatively how one
important characterization of extensive chaos---the
Lyapunov dimension density---depends on spatial
disorder, as measured by the spatial correlation
length~$\xi$. Surprisingly, we find that a
representative extensively chaotic system does not act
dynamically as many weakly interacting regions of
size~$\xi$. }

More specifically, researchers have conjectured
\cite{Cross93} that the fractal dimension~$D$ of a
sufficiently-large homogeneous spatiotemporal chaotic
system should obey the following simple scaling
relation:
\begin{equation}
  D = C (L/\xi)^d \qquad \hbox{for $L \gg \xi$} .
\label{scaling-law}
\end{equation}
Here~$C$ is a constant and $d$~is the spatial
dimensionality of the system, e.g., $d=2$ for a
large-aspect-ratio convection experiment
\cite{Ahlers85}.  \eqref{scaling-law} follows from an
assumption of spatial disorder, that parts of a large
homogeneous system are uncorrelated and hence
dynamically independent when separated by distances
larger than the length~$\xi$.  The volume~$L^d$ of the
system then acts as a gas of $N=L^d/\xi^d$
weakly-interacting regions of volume~$\xi^d$ which
contribute independently to the fractal dimension.

\eqref{scaling-law} predicts that spatiotemporal chaos
is extensive since the fractal dimension~$D$ is
proportional to the system volume~$L^d$. This has
indeed been verified in numerical calculations of
one-dimensional chaotic systems
\cite{Manneville85,Grassberger89,Sirovich90} and is
believed to hold generally.  A more subtle prediction
of \eqref{scaling-law} is that extensive chaos depends
on the spatial disorder through a single
lengthscale~$\xi$ so that~$D \propto 1/\xi^d$ for a
fixed but sufficiently large system size~$L$. It is
useful to formulate this second prediction in terms of
an intensive parameter so that the system size~$L$ and
details of the boundary conditions no longer enter.
This can be done by introducing the intensive dimension
density~$\delta = D / L^d$, which is the fractal
dimension per unit volume for an extensively chaotic
system.  \eqref{scaling-law} then predicts that
\begin{equation}
  \delta \propto 1/\xi^d , \label{delta-vs-xi}
\end{equation}
which we test in detail below. Researchers have shown
for one case that this quantity is independent of the
choice of boundary conditions for sufficiently large
system sizes (see Fig.~10 of
\refcite{Torcini91}).

Eqs.~(\ref{scaling-law}) and~(\ref{delta-vs-xi}) are
important to verify and to understand for two reasons.
First, they suggest a simple theoretical explanation
for why chaos should be extensive, relating the
complexity~$D$ of a dynamical system (roughly the
minimum number of degrees of freedom needed to describe
a system \cite{Parker89}) to its spatial disorder.
Second, \eqref{scaling-law} suggests a straightforward
way to estimate the dynamical complexity of many
interesting experimental systems by simply measuring
the correlation length~$\xi$
\cite{Coullet89a,Bayly93a}.  This approach would have
many benefits compared to time series methods
\cite{Casdagli91} since large spatiotemporal chaotic
systems are high-dimensional, the length of a time
series needed to estimate fractal dimensions above five
is prohibitively large \cite{Ruelle90}, and algorithms
for estimating fractal dimensions from time series are
notoriously unreliable \cite{Casdagli91,Lorenz91}.  As
an example, we note that scientists recently estimated
a correlation length of~$\xi \approx 6 \,\rm mm$ from
voltage measurements on the surface of a fibrillating
pig heart \cite{Bayly93a}.  Since the radius of the
heart was~$R \approx 25 \,\rm mm$, \eqref{scaling-law}
would suggest that the fibrillating heart involved
approximately $D \ge 4 \pi R^2 / \pi \xi^2 \approx 70$
independent regions, a result that is consistent with
the failure of time-series algorithms to detect
low-dimensional chaos \cite{Kaplan90}.

To investigate the dependence of the dimension
density~$\delta$ on the correlation length~$\xi$ we
have studied chaotic numerical solutions of the
one-dimensional complex Ginzburg-Landau equation
\cite{Cross93}:
\begin{equation}
   \partial_t u(x,t) =
  u
  +  (1 + i c_1) \partial^2_x u
  -  (1 - i c_3) |u|^2 u
  , \label{1d-cgl}
\end{equation}
in a periodic system of size~$L$.  The field~$u(x,t)$
is complex-valued while the parameters~$c_1$ and~$c_3$
are real-valued.  \eqref{1d-cgl} is an experimentally
relevant \cite{Glazier91,Ning93b} continuum dynamical
model that holds universally near the onset of a Hopf
bifurcation from a static homogeneous state to an
oscillatory state \cite{Cross93}. It is particularly
valuable for testing \eqref{delta-vs-xi} since there is
an apparent phase transition in large systems
\cite{Shraiman92}---the L1 transition---as the
parameter~$c_3$ is varied for a fixed parameter~$c_1
\ge 1.8$. The correlation length~$\xi$ increases
rapidly as one approaches the L1~transition from one
side and so yields a valuable strong variation of~$\xi$
with parameters.

We integrated \eqref{1d-cgl} using a pseudospectral
method with time-splitting of the operators
\cite{Canuto88}, carefully checking our results for
convergence with respect to spatial and temporal
resolutions and with respect to total integration
times.  Correlation lengths~$\xi$ were obtained by
examining the asymptotic exponential decay $A
\exp(-x/\xi)$ of the magnitude of the complex two-point
equal-time correlation function $C(x) =
\langle u^\ast(x+x',t) u(x',t) \rangle$, where the
angle brackets indicate averaging over time~$t$ and
space~$x'$. We calculated a particular fractal
dimension, the Lyapunov dimension~$D_{\cal L}$, via the
Kaplan-Yorke formula which expresses $D_{\cal L}$ in
terms of the spectrum of Lyapunov exponents
\cite{Parker89}.  The exponents were obtained by an
expensive but widely-used algorithm \cite{Parker89}
that involves integrating \eqref{1d-cgl} together with
many copies of the linearization of this equation
around the solution~$u$.  We used about 1200 CPU hours
on CRAY YMP and on Thinking Machines~CM5 supercomputers
to obtain the results reported below.  For most
dynamical systems, the Lyapunov dimension is believed
to be the same as the information dimension~$D_1$,
which is just one of an infinity of fractal
dimensions~$D_q$ that characterize multifractal strange
attractors \cite{Eckmann85}.

Our results are presented in Figures~1-3 for the fixed
parameter value~$c_1 = 3.5$ and for various values of
the parameter~$c_3$ near the~L1 transition. We have
calculated correlation lengths~$\xi$ for various system
sizes up to~$L=262,144$ and Lyapunov
dimensions~$D_{\cal L}$ for system sizes only up
to~$L=1024$ (since the latter is far more expensive to
calculate numerically).  \Figref{extensive-chaos}(a)
confirms the prediction of \eqref{scaling-law}
concerning extensive chaos by showing that~$D_{\cal L}$
grows linearly with system size~$L$ over the entire
range~$64 \le L \le 1024$; for~$L=1024$, we estimate a
large fractal dimension of~$D \approx 152$. The
dimension density~$\delta$ was estimated by a
least-squares estimate of the slope of the linear
portion of this curve.  Similar extensive chaos curves
were obtained for all~$c_3$ values reported in this
paper.  \Figref{extensive-chaos}(b) shows how we
calculated the correlation length~$\xi$ from the
asymptotic exponential decay of the spatial correlation
function~$C(x)$. Near the L1~transition, an
unexpectedly large system size of~$L / \xi > 1000$ was
needed to find good exponential decay and to find a
value of~$\xi$ that was independent of the system size.

\figref{delta-xi-versus-c3} summarizes how the
Lyapunov dimension density and spatial correlation
length vary with the parameter value~$c_3$ for the
particular value~$c_1=3.5$. As~$c_3$ decreases through
the approximate position of the~L1 transition ($c_3
\approx 0.74$), the Lyapunov dimension density
(\figref{delta-xi-versus-c3}(a)) decreases continuously
within our numerical accuracy.  The correlation
length~$\xi$ has contrary behavior and increases by one
order of magnitude close to the L1~transition (from~6
to~172).  Careful numerical tests suggest that~$\xi$
can not be estimated reliably below~$c_3 \le 0.79$,
probably because it has become too large. As we
describe elsewhere in more detail, we find that the
$\xi$ vs~$c_3$ data are well described by an algebraic
divergence of the form $\xi
\propto \left(c_3 - c_3^\ast\right)^\alpha$ with
exponent~$\alpha=-2.4$ and with~$c_3^\ast = 0.735$.

Our key result is given in
\figref{dimdensity-versus-xi}, which presents the data
of \figref{delta-xi-versus-c3} in a way appropriate for
testing \eqref{delta-vs-xi}. For spatial
dimension~$d=1$, this equation predicts that the
dimension density~$\delta$ should be linear in the
quantity~$1/\xi$ while, in fact, the observed
dependence is unambiguously nonlinear (the inset figure
suggests that a $1/\sqrt{\xi}$ dependence may be more
appropriate). If~$\xi$ is truly diverging across the~L1
transition while~$\delta$ changes continuously,
\eqref{delta-vs-xi} must break down.  This result is
quite surprising since the chaos is extensive
(\figref{extensive-chaos}(a)) and an intuitive reason
for the extensive chaos is that remote parts of the
system are uncorrelated and so contribute independently
to the dynamics.

We conclude that Eqs.~(\ref{scaling-law})
and~(\ref{delta-vs-xi}) are not correct. One
explanation may be that the prefactor~$C$ in
\eqref{scaling-law} is not constant but depends
on~$\xi$ itself, i.e., details of the local dynamics
over a region of size~$\xi$ must be included in the
analysis. Another reason may be that there is more than
one lengthscale needed to characterize Ginzburg-Landau
spatiotemporal chaos. A third possibility is that
Eqs.~(\ref{scaling-law}) and~(\ref{delta-vs-xi}) are,
in fact, correct but involve a new, as-of-yet
unidentified, lengthscale~$\lambda$ in place of the
correlation length~$\xi$ \cite{Cross93,Fraser89}.
Further analysis will be needed to distinguish between
these possibilities and to understand how the dynamical
complexity of an extensively chaotic system is related
to its spatial disorder.

We thank P.~Bayly, H.~Chate, M.~Cross, G.~Grinstein,
E.~Lorenz, J.~Socolar, and L.~Unger for useful
discussions.  This work was supported by the National
Science Foundation, by the Office of Naval Research, by
the North Carolina Supercomputing Center, and by the
National Center for Supercomputing Applications.

\newpage
\begin{thefigures}{99}

\figitem{extensive-chaos}

{\bf (a)} Plot of the Lyapunov fractal
dimension~$D_{\cal L}$ versus system size~$L$ for
\eqref{1d-cgl} with periodic boundary conditions. We
used the parameters values~$c_1=3.5$ and~$c_3=0.8$, a
constant time step of~$\Delta{t} = 0.05$, a spatial
resolution of two Fourier modes per unit length, and a
total integration time of~$T=50,000$ time units.  The
error in each fractal dimension is smaller than the
size of the plotted points.

{\bf (b)} Log-linear plot of the magnitude of the
spatial correlation function $C(x) = \langle
u^\ast(x+x',t) u(x',t) \rangle$ for spatiotemporal
chaotic solutions of \eqref{1d-cgl} for the parameter
values~$c_1=3.5$ and~$c_3=0.8$. From left to right, the
curves correspond to system sizes respectively
of~$L=1024$, $16384$, $32768$, $131072$, and~$262144$.
A clear exponential decay of correlations is found but
only for system sizes that are surprisingly large
compared to the correlation length of~$\xi = 126$. The
space-time resolution and total integration time were
the same as for Part~(a).

\figitem{delta-xi-versus-c3}

{\bf (a)} Plot of the Lyapunov dimension
density~$\delta$ versus the parameter~$c_3$ for
fixed~$c_1=3.5$. Dimension densities were obtained from
the slopes of figures similar to
\figref{extensive-chaos}(a). To within numerical
accuracy, the curve is continuous across the~L1
transition region \cite{Shraiman92}.

{\bf (b)} Plot of the spatial correlation length~$\xi$
versus the parameter~$c_3$ for fixed~$c_1=3.5$. There
is an apparent divergence which is the signature of
long-range order arising from a phase transition. The
$\xi$ values were obtained by least-squares fits to the
linear portion of correlation plots similar to
\figref{extensive-chaos}(b). Systems of increasing
sizes up to~$L=262,144$ were studied until we had found
values of~$\xi$ that no longer changed upon further
increases in~$L$. We were not able to calculate
correlation lengths reliably for~$c_3 < 0.79$ since the
value of~$\xi$ continued to change with increasing
system size and with increasing integration time.

\figitem{dimdensity-versus-xi}

Plot of the Lyapunov dimension density $\delta$ versus
inverse correlation length~$1/\xi$ for the parameter
value~$c_1=3.5$. The inset shows the same data plotted
versus the inverse square root, $1/\sqrt{\xi}$.  The
dotted lines connecting adjacent points were added only
to guide the eye in identifying possible linear trends.
The nonlinear dependence on~$1/\xi$ is inconsistent
with the scaling law \eqref{scaling-law} for a gas of
weakly coupled domains that are highly correlated over
the range of the correlation length~$\xi$.

\end{thefigures}

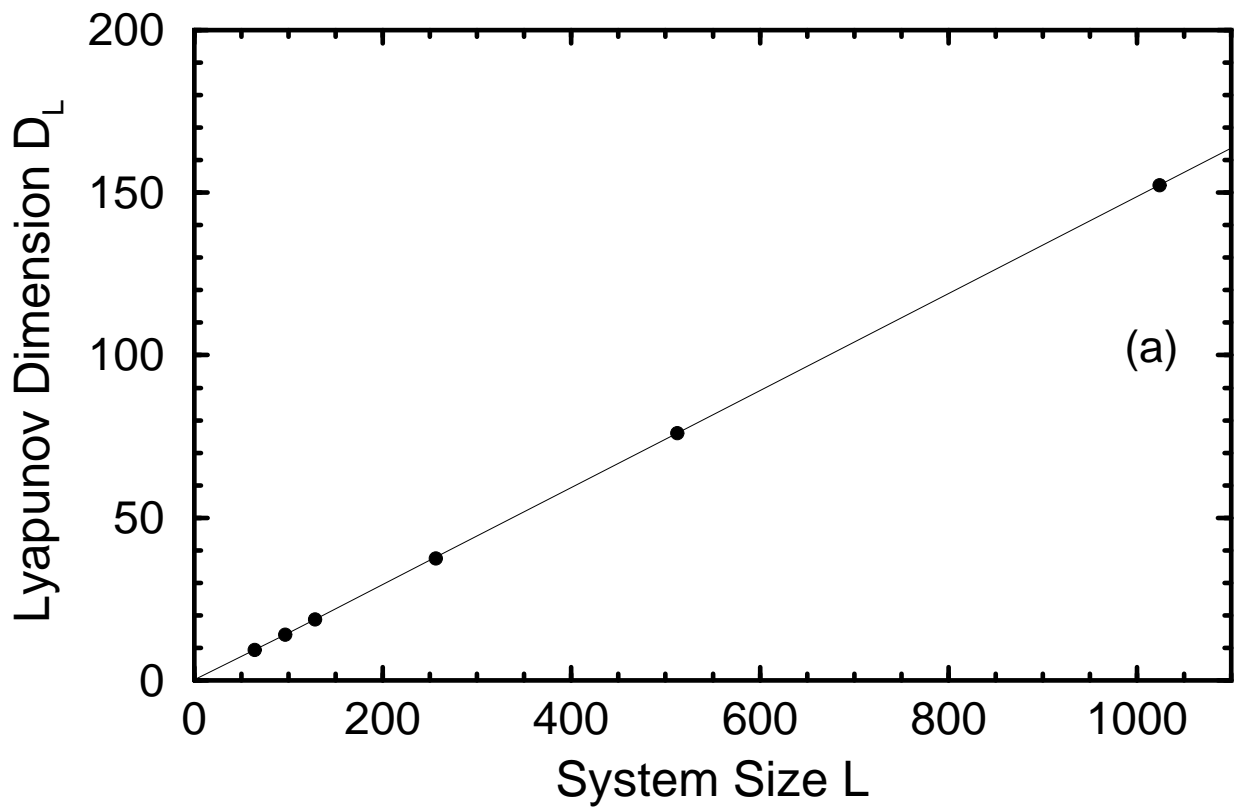
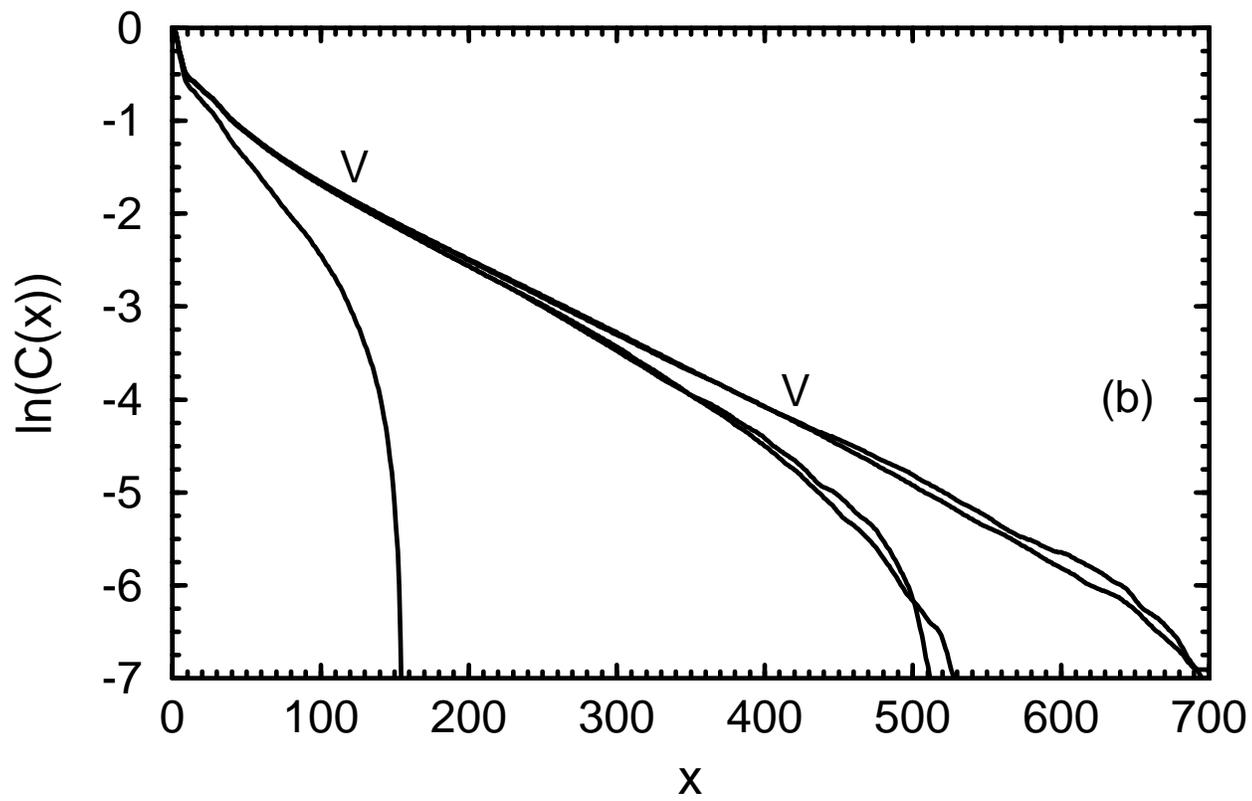
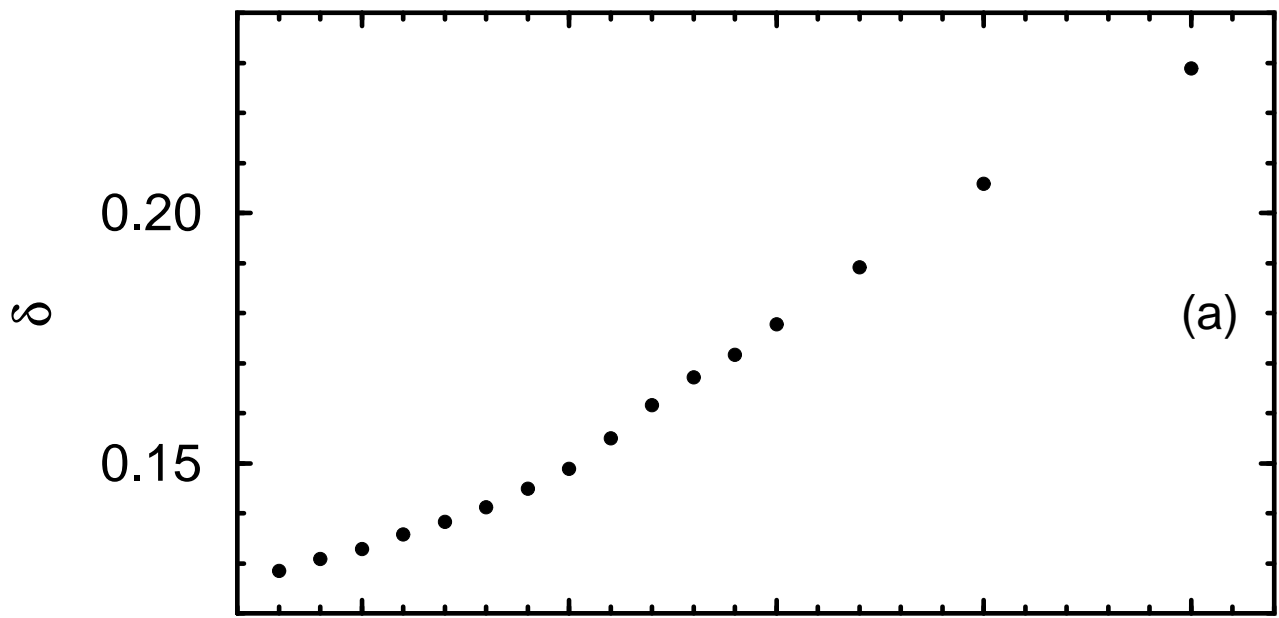
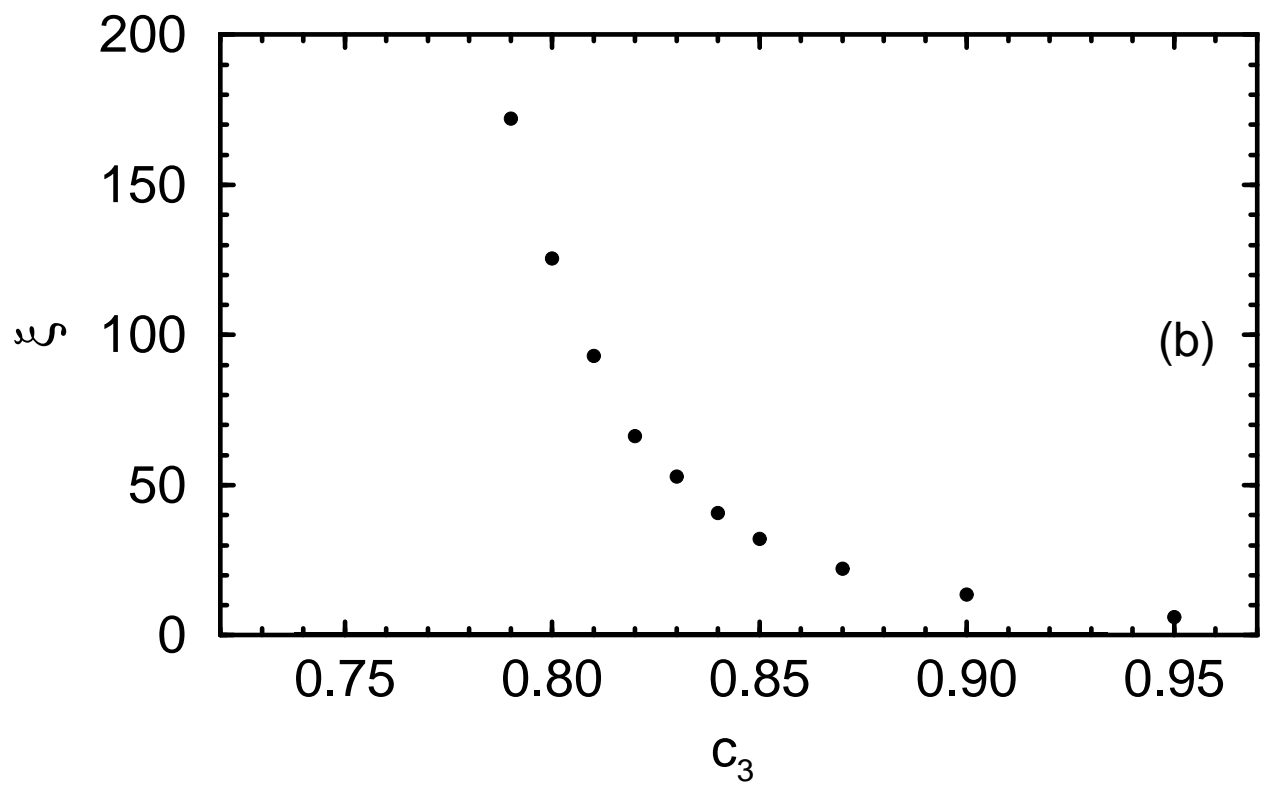
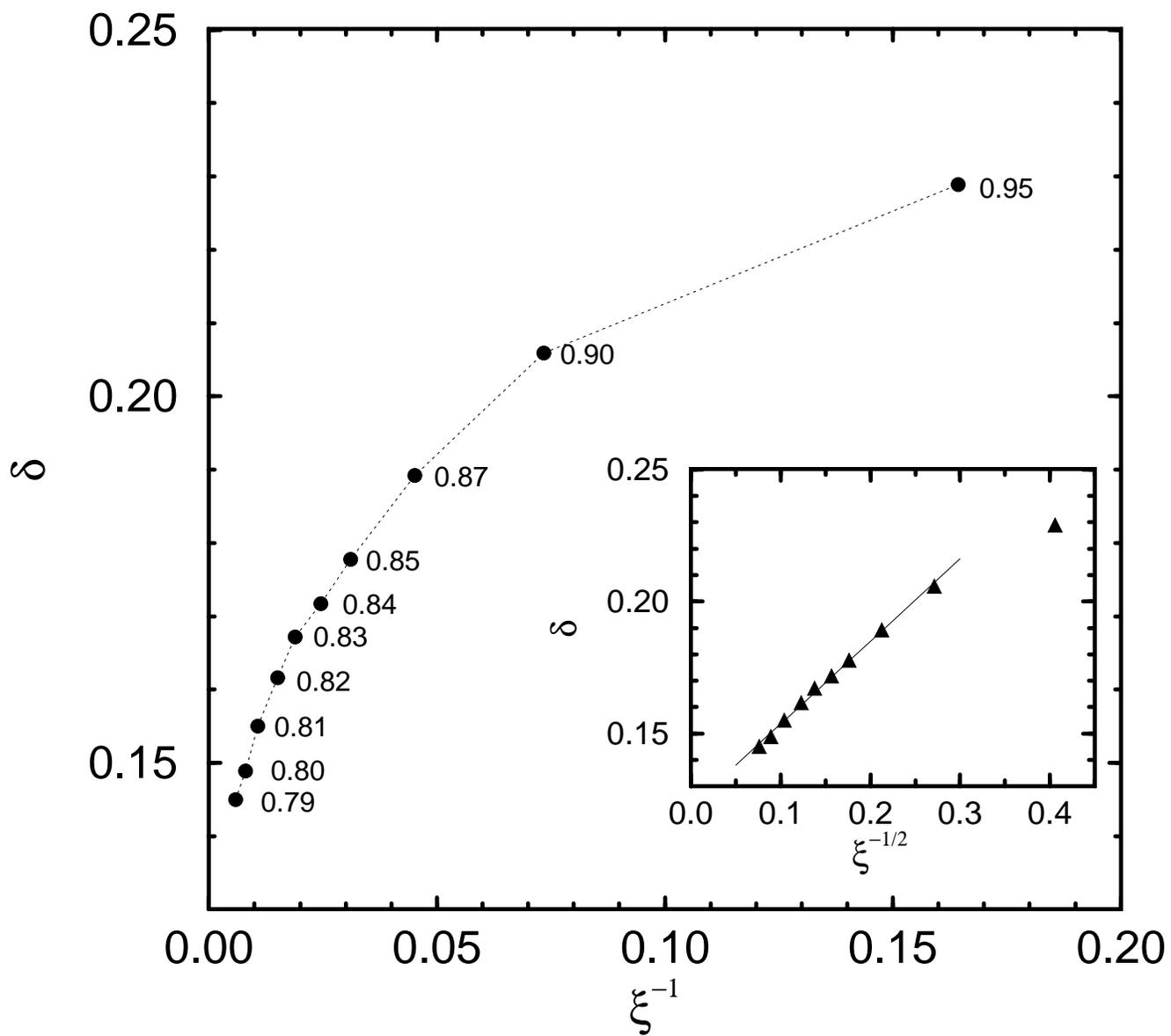
\end{document}